\documentclass[12pt]{article}
\usepackage{calrsfs}
\usepackage{ucs}
\usepackage[utf8x]{inputenc}
\usepackage{braket}
\usepackage{dsfont}
\usepackage[colorlinks=true, citecolor = red, linkcolor = blue, urlcolor  = red]{hyperref}
\usepackage[dvipsnames]{xcolor}
\usepackage{amsmath,amssymb}
\def\a{\alpha}
\def\b{\beta}

\def\d{\delta}
\def\e{\epsilon}

\def\k{\kappa}

\def\m{\mu}

\def\t{\theta}

\def\be{\begin{equation}}
\def\ee{\end{equation}}
\def\arr{\begin{array}{rll}}
\def\ea{\end{array}}
\def\bea{\begin{eqnarray}}
\def\eea{\end{eqnarray}}
\def\cN{{\cal{N}}}

\def\N2{$N{=}2$}

\def\>{\rangle}
\def\<{\langle}
\def\+{\dagger}
\def\={\ =\ }

\begin{document}
\vspace{0.5cm}
\renewcommand{\thefootnote}{\fnsymbol{footnote}}
\renewcommand{\thefootnote}{\fnsymbol{footnote}}
\begin{titlepage}
\setcounter{page}{0}
\vskip 1cm
\begin{center}
{\LARGE\bf On $OSp(N|2)$ superconformal mechanics }\\

\vskip 1cm
$
\textrm{\Large Dmitry Chernyavsky \ }
$
\vskip 0.7cm
{\it
School of Physics, Tomsk Polytechnic University, \\
634050 Tomsk, Lenin Ave. 30, Russian Federation} \\
{E-mail: chernyavsky@tpu.ru}

\end{center}
\vskip 1cm
\begin{abstract} \noindent
Superparticle models with $OSp(N|2)$ supersymmetry group are studied. We first consider the $N=4$ case and construct the models with $\k$--symmetry on the coset spaces of the $OSp(N|2)$ supergroup. In addition, within the canonical formalism we present an $OSp(4|2)$ superparticle model with semi--dynamical angular variables. For generic $N$ we construct a superparticle model on $AdS_2\times S^{N-1}$ with the reduced $\k$--symmetry. It is demonstrated that the Hamiltonian of this model has the same structure as the one for the $N=4$ case because additional fermions contribute to the second--class constraints only.
\end{abstract}

\vskip 1,5cm


\end{titlepage}

\renewcommand{\thefootnote}{\arabic{footnote}}
\setcounter{footnote}0

\noindent

\section{Introduction}
One--dimensional superconformal systems can be thought of as the limiting case of the higher--dimensional analogues which capture silent features of the latter in a simpler context. Typically one is concerned with one--dimensional (super)conformal mechanics because they provide dual theories for two--dimensional gravity \cite{Klemm_1, Klemm_2, Jak, Vass, Vass_2}. On the other hand, they are of interest on their own. For instance, some $d=1$ supermultiplets cannot be obtained by dimensional reduction from higher dimensions.

$d=1$, $\mathcal{N}=4$ superconformal algebras and their mechanical realizations attracted particular interest due to the proposal that they might be relevant for a microscopic description of extreme black holes \cite{GT, Kallosh}. This proposal inspired extensive studies \cite{Wyllard}-\cite{Gal_D21}, which proved useful for understanding the structure of interactions of $\mathcal{N}=4$ supermultiplets of $su(1,1|2)$, $osp(4|2)$ and $D(2,1;\a)$. Some of these models are described by the $\k$--symmetric actions \cite{Zhou, Anabalon_Zanelli, Ch_2}.


It is natural to wonder what happens beyond the $\mathcal{N}=4$ case. The most powerful approach for constructing the superconformal mechanics is the superfield formalism. However, in practical applications it may turn out to be difficult to single out irreducible supermultiplets by imposing constraints on superfields \cite{ABC}. Another way is the direct construction of the superconformal mechanics within the canonical formalism \cite{Galajinsky_Superparticle,Gal_SU(2),Gal_D21,GL}. This method relies upon the structure relations of the superalgebra, and an going beyond $\mathcal{N}=4$ may turn out to be problematic \cite{GL}\footnote{However, see a recent paper \cite{KLS}, where $\cN$--extended Calogero model with spin variables was constructed.}. Superconformal mechanics can also be constructed by higher--dimensional reduction \cite{Okazaki,Copland}. Finally, by using the method of nonlinear realizations one can directly build the action, whose form can be restricted by requiring the $\k$--symmetry \cite{Zhou,Zhou_2,Anabalon_Zanelli,Ch,Ch_2,Heinze_1,Heinze_2}.
 In a recent paper \cite{Ch_2}, this method was applied for the construction of superpaticle models on the coset spaces of the $SU(1,1|N)$ supergroup. It was shown that the $\k$--symmetry for the general $N$--case is broken down to a one--parametric fermionic gauge symmetry. This work gives an explanation for the problem encountered in \cite{GL}, in which the authors vainly attempted to construct an $SU(1,1|N)$ superparticle system with angular variables by a straightforward generalization of the $SU(1,1|2)$ case.

In the present paper, we extend the analysis of \cite{Ch_2} to the $osp(N|2)$ superalgebra. The isntance $N=4$ is special. We show that it is the only nontrivial case when one is able to construct an action on the corresponding coset space with unbroken $\k$--symmetry. Remarkably enough, the coset of $OSp(4|2)$ supergroup with the bosonic part $AdS_2\times S^2$ is also of particular interest from the superfield standpoint, as it represents the only superextension of $AdS_2\times S^2$ which admits a superconformally flat supervielbein and superconnections \cite{FIL}. For generic $N$ the $\k$--symmetry is broken down to a one--parametric gauge symmetry in analogy with \cite{Ch_2}.  A remarkable fact, however, is that the Hamiltonian structure remains the same as in the $N=4$ case, because the additional fermionic degrees of freedom show up in second class constraints only.

The organization of the work is the following. We remind the structure relations of the $osp(N|2)$ superlagebra in the next section. In sect. 3 we investigate the $N=4$ case and construct models on two different coset spaces reproducing some of the models in \cite{Gal_D21}. In addition, we construct an analogue of the model \cite{FIL_2,FIL} with semi--dynamical angular degrees of freedom. Sect. 4 is devoted to the general $N$ case, where we construct the action with the reduced $\k$--symmetry and analyze the model within the canonical formalism. We conclude and summarize in sect. 5. Appendices contain some technical details regarding the explicit form of the Maurer--Cartan one--forms.
\section{Superalgebra}

We start with the standard form of the $osp(N|2)$ superalgebra:
\begin{align}
&
[ H,D ]=H\ , && [ H,K ]=2D\ ,
\nonumber\\[2pt]
&
[D,K]=K\ , && [ J^a,J^b ]=f^{abc} J^c\ ,
\nonumber\\[2pt]
& [ D,Q_j] = -\frac{1}{2} Q_j\ , && [ D,S_j] =\frac{1}{2} S_j\ ,
\nonumber\\[2pt]
&
[ K,Q_j ] =S_j, && [ H,S_j ]=-Q_j\ ,
\nonumber\\[2pt]
&
[ J^a,Q_j] =-\lambda^a_{jk} Q_k\ , && [ J^a,S_j] =-\lambda^a_{jk} S_k\  ,
\nonumber\\[2pt]
&
\{Q_j,Q_k\}=-2iH \d_{jk},\ && \{S_j,S_k\}=-2iK \d_{jk}\,
\nonumber\\[2pt]
&
\{Q_j,S_k\}=2iD\d_{ij}+i \lambda^a_{jk} J^a,
\end{align}
where condensed notation was introduces for the $so(N)$ generators $J^{jk}\rightarrow J^a$ satisfying the algebra
\be\label{so(N)}
[J^{ij},J^{kl}]=\d^{jk}J^{il}+(3\ terms).
\ee
All generators are real.

The bosonic subalgebra is the direct sum $so(1,2)\oplus so(N)$, where the $d=1$ conformal algebra $so(1,2)$ is presented by $\{H, K, D\}$. Fermionic part of the superalgebra is generated by the real supersymmetry charge $Q_i$ and the superconformal partner $S_i$ with $i=,1\dots,N$. The antisymmetric matrices $\lambda^{ij}_{kl}\rightarrow \lambda^a_{kl}$ define $N$--dimensional representation of the rotation algebra
\be
[\lambda^{a},\lambda^b]=f^{abc}\lambda^c,
\ee
where $f^{abc}$ are the $so(N)$ structure constants (\ref{so(N)}). These matrices are supposed to satisfy a Fierz--like identity
\be
\lambda^a_{ij}\lambda^a_{kl}=\d_{ik}\d_{jl}-\d_{il}\d_{jk}.
\ee
For convenience we make a standard decomposition of the $so(N)$ generators into $so(N-1)$ subalgebra and the set of operators, which will be used to generate the cosets
\be\label{algebra decomposition}
J^{mn} := M^{mn}, \qquad J^{mN} := P^m, \qquad m,n=1,\dots, N-1.
\ee
In these notations commutation relations of $so(N)$ read
\bea\label{Algebra_Decomposition}
&&
[M^{mn},P^q]=\d^{qm} P^n-\d^{qn} P^m, \qquad [P^m,P^n]=M^{mn},
\nonumber\\[2pt]
&&
[M^{mn},M^{pq}]=M^{mp}\d_{nq}+(3\ terms).
\eea

We define the Maurer--Cartan (MC) one--forms for a coset element $\tilde G$ in the standard way
\be
\tilde{G}^{-1} d \tilde G=H L_H + K L_K+D L_D+L^a J^a+L_Q Q+L_S S,
\ee
where summation over the rotation indices carried by fermions is assumed. In what follows, we will need the bosonic one--forms only. When analyzing the $\k$--symmetry properties of the actions, the special form of variations of these one--forms will be needed, which can be retrieved from the MC equations (see e.g. \cite{MC_2})
\bea\label{MC variations}
&&
\d L_H=d[\d x_H] +[\d x_D]L_H-[\d x_H]L_D+2i[\d \psi]L_Q,
\nonumber\\[2pt]
&&
\d L_K=d[\d x_K] -[\d x_D]L_K+[\d x_K]L_D+2i[\d \eta]L_S,
\nonumber\\[2pt]
&&
\d L_D=d[\d x_D]-[\d x_H]L_K+2[\d x_K]L_H-2i\left([\d \psi]L_S+[\d\eta]L_Q\right),
\nonumber\\[2pt]
&&
\d L^a=d[\d x^a]-f^{abc}[\d x^b]L^c-i [\d \psi]\lambda^a L_S+ i L_Q\lambda^a[\d \eta],
\eea
where we introduced the notation
\be
[\d Z^A]=L^A{}_M\d Z^M,
\ee
for a MC one--form $L^A=L^A{}_M dZ^M$.
The $\k$--symmetry is characterized by the vanishing of the bosonic variations on the coset with respect to \cite{MC_2} (see also \cite{Ch})
\be\label{Variations_Bosonic}
[\d_\k x_H]=[\d_\k x_K]=[\d_\k x_m]=0.
\ee

\section{Special case: $N=4$}
Before turning to generic $N$, we discuss the instance of $N=4$. In this section we reproduce the known models of $OSp(4|2)$ superconformal mechanics and construct some novel ones.

\subsection{Internal degrees of freedom}

Technically the issue of finding a $\kappa$--symmetric action amounts to using the properties of the matrices $\lambda^m$ corresponding to the coset generators $P_m$, which depend on the way the subalgebra $SO(N-1)$ was extracted from $SO(N)$. In this sense the case of $N=4$ is subtle, because the algebra $so(4)$ is the direct sum of two copies of $so(3)$
\be\label{SO(4)}
[P_{\pm}^m, P_{\pm}^n]=\d_{\pm,\pm} \epsilon^{mnq}P_{\pm}^q.
\ee
As the superalgebra $osp(4|2)$ involves two copies of $so(3)$ in a symmetric manner, the coset space can be generated by any of these. In accordance with the structure relations of $so(4)$, we introduce a set of matrices
\bea\label{so(4)  matrices}
&&
\lambda_+^{1}=\mathds{1}\otimes i \sigma_2, \qquad \lambda_+^{2}=i \sigma_2\otimes  \sigma_1, \qquad \lambda_+^{3}=i \sigma_2\otimes  \sigma_3,
 \nonumber\\[2pt]
&&
\lambda_-^{1}=i \sigma_2\otimes \mathds{1}, \qquad \lambda_-^{2}= \sigma_1\otimes  i \sigma_2, \qquad \lambda_-^{3}= \sigma_3\otimes i \sigma_2,
\eea
which satisfy the (anti)commutation relations
\be\label{Matrix_Anti_Commutator}
[\lambda^{m}_\pm,\lambda^{n}_\pm]=2\d_{\pm\pm}\e^{mnp}\lambda^{p}_\pm, \qquad \{\lambda^{m}_+,\lambda^{n}_+\}=-2\d^{mn}, \qquad \{\lambda^{m}_-,\lambda^{n}_-\}=-2\d^{mn},
\ee
and the Fierz--like identities
\be\label{Fierz}
(\lambda^{m}_+)_{jk}(\lambda^{m}_+)_{pq}=\d_{jp}\d_{kq}-\d_{jq} \d_{pk}+\e_{jkpq}, \qquad (\lambda^{m}_-)_{jk}(\lambda^{m}_-)_{pq}=\d_{jp}\d_{kq}-\d_{jq} \d_{pk}-\e_{jkpq}.
\ee
Here $\sigma_i$ are the Pauli matrices.

\subsubsection{Angular variables parametrizing $S^3$}
Now we turn to the construction of the coset on which an invariant action can be build. The building blocks for the action are the MC one--forms and their invariant quadratic combinations. First of all, let us note that the stability subgroup is allowed to be generated by the bosonic operators only. It proves to be impossible to construct quadratic combinations of the MC one--forms invariant under the left action of the group otherwise. In addition, the $\kappa$--symmetry also imposes strong constraints on the structure of the action. One may see that the coset of a maximal dimension which allows one to construct a $\k$--invariant supersymmetric action is $OSp(4|2)/SO(3)$. The bosonic part of this space is $AdS_3 \times S^3$. As it has a larger supersymmetry group, we refrain from considering it in this work (see e.g. \cite{MC}). The next case is
\be\label{Coset}
\frac{OSp(4|2)}{SO(1,1)\times SO(3)} \sim \frac{\{H, K, D, P_{\pm}^m, Q, S \}}{\{D, P_-^m\}}
\ee
with the bosonic part $AdS_2\times S^3$. We construct the action on this coset space
\be\label{Action_S3}
S=-2m\int \sqrt{L_H L_K- L^n_+ L^n_+}-a\int L_D,
\ee
where the MC one--forms $L_+^m$ correspond to the generators $P_+^m$ and $a$, $m$ are constant parameters. In the pure bosonic background the first term under the root represents a metric on  $AdS_2$, while the second term corresponds to $S^3$. As the stability subgroup is generated by the bosonic operators only, the invariants built in terms of the MC one--forms in the pure bosonic case, are invariants of the whole group as well. Linear term proportional to $L_D$ is transformed as an Abelian connection (see e.g. \cite{Krivonos_lectures}) and can be thought of as the Wess--Zumino term \cite{Ch}. Taking into account these observations, one concludes that the action (\ref{Action_S3}) is invariant under the $Osp(4|2)$ supergroup.

 To see that the action is also invariant under the $\k$--symmetry, we use the standard method \cite{MC_2} which relies upon a technically convenient representation for variations of the MC one-forms. For the case at hand they are given in (\ref{MC variations}) and should be supplemented by (\ref{Variations_Bosonic}). Taking it into account, one can write down a variation of the
action (\ref{Action_S3}) irrespective of a specific choice of the coset parametrization
 \bea\label{ActionVariation}
 &&
 \delta_\k S=-2i\int \left\{\frac{ m L_H [\d_\k \eta]-  m L_+^m [\d_\k\psi] \lambda^m_+}{\sqrt{L_HL_K-L_+^m L_+^m}}- a  [\d_\k\psi] \right\}L_S
 \nonumber\\[2pt]
&&
\qquad \qquad \qquad \qquad  -2i\int \left\{\frac{ m L_K [\d_\k \psi]+  m L_+^m [\d_\k\eta] \lambda^m_+}{\sqrt{L_HL_K-L_+^m L_+^m}}- a [\d_\k\eta] \right\}L_Q,
 \eea
 where the boundary term $[\d_\k x_D]$ coming from the variation of the Wess--Zumino term $L_D$ was discarded. Demanding (\ref{ActionVariation}) to vanish, one obtains a system of linear algebraic equations on $[\d_\k\psi]$ and $[\d_\k\eta]$.
 A solution exists for an arbitrary $[\d_\k\eta]$ (or $[\d_\k\psi]$),
 provided the parameters are subject to the constraint
 \be\label{Parameter restriction}
 m=|a|.
 \ee
For definiteness we may set
\be
[\d_\k \psi]= \frac{\sqrt{L_HL_K-L_+^m L_+^m}}{L_K} \kappa + L^m_+\kappa\lambda_+^m, \qquad [\d_\k\eta]=\kappa,
\ee
 where $\k=\k_i$ with $i=1,\dots,4$, is an arbitrary Grassmann-valued parameter.

Let us now construct the action explicitly. The gauge--fixed MC one--forms on the coset space (\ref{Coset}) are given in Appendix A.
Making the change of the coordinates
\be
t\rightarrow t+\frac{1}{z},\qquad \psi\rightarrow \frac{\psi}{z},
\ee
one brings the action (\ref{Action_S3}) to the form
\be
S=-2m\int \left[z^2-\dot{z}-i\psi\dot{\psi}-e^m e^m+ie^m(\psi \lambda_+^p\psi)O^{mp}+\frac{1}{4}\e_{ijkl} \psi_i\psi_j\psi_k\psi_l\right]^{1/2}dt+2m\int z dt.
\ee
 To establish a relation between our models and those constructed previously, we turn to the canonical formalism. Introducing the momenta $(p_z, p_\t, p_\phi, p_\chi)$, one finds the Hamiltonian
\be\label{Hamiltonian_3}
H=z^2 p_z+\frac{m^2}{p_z}+2mz+\frac{1}{4 p_z}J^m_+ J^m_+-\frac{i}{2}J^m_+ (\psi \lambda^m\psi),
\ee
where
\bea\label{so(3) generators}
&&
J^1_+=-p_\phi \cot\t \cos\phi-p_\t\sin\phi+p_\chi \cos\phi \sin^{-1}\t,
 \nonumber\\[2pt]
&&
J^2_+=-p_\phi \cot\t \sin\phi+p_\t\cos\phi+p_\chi \sin\phi \sin^{-1}\t, \qquad J^3_+=p_\phi,
\eea
are the bosonic parts of the rotation generators $P^m_+=J^m_++\frac12 (\psi\lambda_+^m\psi)$.
Fermionic momenta give rise to the second class constraints
\be
p_\psi-ip_z \psi=0.
\ee
To put the Hamiltonian into the standard form, one makes the canonical transformation
\be\label{Canonical redefinition}
z\rightarrow -\frac{p}{x}-\frac{2m}{x^2}, \qquad p_z\rightarrow \frac{x^2}{2}, \qquad \psi\rightarrow \frac{\psi}{x}, \qquad p_\psi\rightarrow x p_\psi,
\ee
which leads to
\be\label{Hamiltonian}
H=\frac{p^2}{2}+\frac{1}{2x^2}J^m_+J^m_+-\frac{i}{2x^2}(\psi\lambda^m\psi)J^m_+, \qquad p_\psi-\frac{i}{2}\psi=0.
\ee
As was shown above, one copy of $so(3)$ is realized on the bosonic and fermionic variables by $P_+^m$, while the second one acts upon the fermionic variables only $P_-^m=\frac12 (\psi\lambda^m_-\psi)$.  This model is invariant under the additional symmetry transformations, originating from the right action of $SO(3)$ on itself
\bea
&&
J^1_-=\cot\t \cos\chi p_\psi+\sin\chi p_\t+\cos\chi\sin^{-1}\t p_\phi,
 \nonumber\\[2pt]
&&
J^2_-=-\cot\t \sin\chi p_\chi+\cos\chi p_\t-\sin\chi\sin^{-1}\t p_\phi, \qquad J^3_-=p_\chi.
\eea

Summarizing the discussion above, the model is invariant under $OSp(4|2)\times SO(3)$--transformations. At the Hamiltonian level, it is easy to see that the functions $J_+^m$ can be arbitrary generators of the rotation algebra \cite{Gal_SU(2),Gal_D21}. In particular, we may set $p_\chi$ to be a constant, thus breaking the symmetry group down to $OSp(4|2)$ supergroup. In this case the model describes an $OSp(4|2)$ supersymmetric extension of the particle propagating in the near horizon region of an electrically and magnetically charged black hole. Taking into account the isomorphism $osp(4|2) \simeq D(2,1;-\frac12)$, one may verify that the Hamiltonian (\ref{Hamiltonian}) exactly reproduces $D(2,1;\a)$ superconformal mechanics studied in \cite{Gal_D21} for $\a=-1/2$.

 \subsubsection{$S^2$ and magnetic monopole}

The next coset we consider is
 \be\label{Coset_2}
 \frac{OSp(4|2)}{SO(1,1)\times SO(3)\times SO(2)}\sim  \frac{\{H, K, D, P_{\pm}^m, Q, S \}}{\{D, P_-^m, P_+^3\}}.
 \ee
We introduce a superparticle action on this coset by
\be\label{Action_S2}
S=-2m\int \sqrt{L_H L_K-L^m_+ L^m_+}-\int (a L_D+bL_+^3),
\ee
where the summation over $m=1,2$ is assumed. The bosonic part of this action describes a particle on the $AdS_2 \times S^2$ spacetime with the two--from flux and the corresponding background can be thought of as the near horizon region of a nonrotating black hole, where the parameters $a$ and $b$  are proportional to electric and magnetic charges of the black hole, respectively. It represents an $OSp(4|2)$ counterpart of the $SU(1|2)$ superparticle of \cite{Zhou, Zhou_2}. Due to the isomorphism $osp(4|2) \simeq D(2,1;-\frac12)$, the action (\ref{Action_S2}) describes exactly the $\k$--symmetric model constructed in \cite{Ch_2} on the coset space of the $D(2,1;\a)$ group for $\a=-1/2$.

As $S^3$ is a bundle over $S^2$ with $U(1)$ fiber, the action (\ref{Action_S2}) can be obtained from (\ref{Action_S3}) by the dimensional reduction. At the classical level the only effect of this reduction is the fixation of the momentum conjugate to the coordinate parametrizing $U(1)$, thus producing the model of the previous section with constant $p_\chi$. However, the requirement that the action (\ref{Action_S2}) is invariant under the $\k$--symmetry transformations yileds the constraint
 \be\label{Restriction}
m^2=a^2+\frac{b^2}{4}.
\ee
Note that this expression explicitly involves the magnetic charge $b$, which is the momentum $p_\chi$ fixed by the dimensional reduction. Hence, it differs from the model in the previous section with constant $p_\chi$ and (\ref{Action_S2}).


\subsection{Constrained angular sector}
This section we consider the model with constrained angular sector, which is governed by the action
\be\label{Action_S4}
S=-2m\int \sqrt{L_H L_K} +\int\left(b\sqrt{L_+^n L^n_+} - a L_D\right),
\ee
where $n=1,2$ for the coset (\ref{Coset_2}) and  $n=1, 2 ,3$ for (\ref{Coset}). One can see that this action is invariant under the $\k$--symmetry transformations provided the parameters $a$, $b$, $m$ are subject to the constraint (\ref{Restriction}).
In addition to the $\k$--symmetry this action enjoys the reparametrization invariance. Using explicit form of the MC one--forms and imposing the static gauge as in the previous section, one finds the Hamiltonian
\be
H=z^2 p_z+\frac{m^2}{p_z}+2mz-\frac{i}{2}J^m_+ (\psi \lambda^m\psi),
\ee
and the first--class constraint
\be\label{Constraint_1}
J^m_+ J^m_+=b^2,
\ee
where $J_+^m$ are given in (\ref{so(3) generators}) for the coset (\ref{Coset}), while for the coset (\ref{Coset_2}) are the same but with $p_\chi=0$. Implementing the canonical transformation (\ref{Canonical redefinition}) results in the Hamiltonian (\ref{Hamiltonian}) on the constraint surface (\ref{Constraints_1}). One recognizes in this model an $OSp(4|2)$ analogue of the $SU(1,1|2)$ superparticle constructed in \cite{Anabalon_Zanelli}. Note also that this model could be constructed starting from (\ref{Hamiltonian}) and using a projection on the constraint surface (\ref{Constraints_1}). However, there is a subtlety at the quantum level because $b^2$ should be quantized. In contrast to the model obtained by the projection of (\ref{Hamiltonian}), the requirement for the original action (\ref{Action_S4}) to enjoy the $\kappa$--symmetry leads to the relation (\ref{Restriction}),
 thus providing quantization of the mass parameter $m$.

\subsubsection{Semi--dynamical angular variables}

Although the model in the previous section has the same symmetry group and the same number of physical degrees of freedom as the
$OSp(4|2)$--model with spin variables in \cite{FIL} (see also \cite{FIL_2, BK, FIL_4}), its physical content is different.
In this section we pay special attention to the semi--dynamical angular variables.

Let us introduce a set of real variables $\a_i$, $\b_i$, $i=1,\dots,4$ which obey the bracket
\be\label{bracket}
\{\a_i,\b_j\}=-\d_{ij}.
\ee
One can use the variables in order to construct the $so(4)$--generators
\be
J^m_\pm := \left(\a\lambda^m_\pm\b\right).
\ee
Using the Fierz--like identities (\ref{Fierz}) for the matrices $\lambda^m_+$, we can find the Casimir element
\be
J^m_+J^m_+=\a^2 \b^2-(\a_i\b_i)^2.
\ee
To construct an analogue of the superparticle \cite{FIL}, it seems natural to restrict the phase space variables to the surface
\be\label{Constraints_1}
\a^2 \b^2=c_1^2,\qquad \a_i\b_i=c_2,
\ee
provided that they commute with the generators $J^m_\pm$.
These are the first class constraints. They eliminate $(2+2)$ degrees of freedom leaving four physical degrees of freedom. Next, we can define the supercharges
\be
Q_i=p \psi_i-\frac{1}{x}J^m_+(\lambda_+^m \psi)_i ,
\ee
which give rise to the Hamiltonian on the constraint surface
\be\label{Hamiltonian_1}
H=\frac{p^2}{2}+\frac{1}{2x^2}(c^2_1-c^2_2)-\frac{i}{2x^2}(\psi\lambda^m_+\psi)J^m_+.
\ee
At the Lagrangian level the bracket (\ref{bracket}) and the constraints  (\ref{Constraints_1}) can be presented by the term
\be
\frac{1}{2}(\dot{\a}_i\b_i-\a_i\dot{\b}_i)+A(\a^2\b^2-c_1^2)+B(\a_i\b_i-c_2),
\ee
where $A$ and $B$ are the Lagrange multipliers. The kinetic term is of the first order giving rise to the Dirac bracket (\ref{bracket}). Clearly, the model can be considered without a projection on the constraint surface, thus having eight phase space degrees of freedom on which the algebra $so(4)$ is realized. Finally, it should be noted that the model at hand can be straightforwardly generalized to accommodate the $D(2,1;\a)$ symmetry group \cite{Gal_D21}.

\section{The general case}

We now turn to the general case of $osp(N|2)$ superalgebras and their superparticle realizations with the $\k$--symmetry. As will be demonstrated below, the $\k$--symmetry is broken down to the one--parametric fermionic gauge symmetry similar to the $SU(1,1|N)$ superparticle \cite{Ch_2}. We construct our model on the coset
\be\label{Coset_N}
\frac{OSp(N,2)}{SO(1,1)\times SO(N-1)}\sim \frac{\{H, K, D, J^a, Q, S\}}{\{D, M^{mn}\}},
\ee
whose bosonic part is $AdS_2\times S^{N-1}$. In what follows we consider the action (\ref{Action_S3}) but now with $n=1,\dots,N-1$. The issue of the $\k$--symmetry can be addressed within the approach used in the previous sections. The algebraic equations which come from the requirement that the
action is invariant under the $\k$--transformations coincide with those from (\ref{ActionVariation}) but again $n=1,\dots,N-1$. These equations can be put in the form
\be\label{LinearEquation_2}
[\d_\kappa\eta]\left(L^m L^n \{\lambda^m, \lambda^n\}+2 L^m L^n\right)=0,
\ee
provided the relation (\ref{Parameter restriction}) holds, where the curly brackets denote the matrix anticommutator. To compute it explicitly, it is convenient to define the matrices $\lambda^{jk}$ in bra--ket notations
\be
\lambda^{jk}=\ket{j}\bra{k}-\ket{k}\bra{j}, \qquad j,k=1,\dots, N.
\ee
Now, it is easy to find the anticommutator encountered above
\be\label{Anticommutator}
\{\lambda^m,\lambda^n\}=-(\ket{m}\bra{n}+\ket{n}\bra{m})-2\ket{N}\bra{N},
\ee
which implies the one--parametric solution to the equations (\ref{LinearEquation_2})
\be\label{k_solution}
\ket{[\d_\kappa\eta]}=\kappa \ket{N},
\ee
where $\kappa$ is an arbitrary Grassmann--valued parameter. Note that the equations (\ref{LinearEquation_2}) hold for any $[\d_\kappa\eta]$ only in trivial case $N=2$. As it was explained in Sect. 2, the case $N=4$ is special because the algebra $so(4)$ is the direct sum of two copies of $so(3)$ and can be presented by the matrices (\ref{so(4)  matrices}) with the properties (\ref{Matrix_Anti_Commutator}) and (\ref{Fierz}). In the general case, the only decomposition of the $so(N)$ algebra on $so(N-1)$ and the set of operators generating coset is given by  (\ref{algebra decomposition}), which leads to the anticommutation property (\ref{Anticommutator}) and to the result (\ref{k_solution}).

Let us now construct the action explicitly. Taking into account (\ref{MC forms N}), the action can be written in the form
\be
S=-2m\int dt\sqrt{ L_H (z^2L_H+\dot{z} - 2iz (\dot{\psi}\eta)-i (\eta\dot{\eta}))- L^m L^m}-2m \int (z L_H-i\dot{\psi}\eta)dt,
\ee
where all the differentials in $L_H$ and $L^m$ are replaced by the velocities with respect to $t$.
 We may fix the gauge by setting the N'th component of $\eta$ to be vanishing, but it does not lead to simplification of the action. Note that the fermionic contribute quartic terms, which is the general feature of the superparticle and superstring actions constructed with the use of the superalgebra in the standard superconformal basis \cite{Sorokin}.

Let us analyze the model in the canonical formalism. A lengthy but straightforward computation gives the Hamiltonian
 \be
H=\frac{m^2}{p^2_z}+z^2 p_z+2mz+\frac{1}{4p_z} J^a J^a-\frac{i}{2}J^a (\eta\lambda^a\eta),
\ee
where we have used the fact that $e^\m_m O_{am}$ is a set of Killing vectors (see e.g. \cite{Ortin}) and thus $p_\m e^\m_m O_{am}:=J^a$ is a set of phase space generators of $so(N)$ algebra. Fermionic momenta give rise to the constraints
\bea
&&
p_\eta-ip_z\eta=0, \\\label{Constraint_Additional}
&&
p_\psi+i\psi \left(\frac{m^2}{p_z}-\frac{1}{p_z} J^a J^a+z^2 p_z+2mz\right)-2zp_\eta+i J^a(\eta \lambda^a)+im\eta=0.
\eea
 Note that the Hamiltonian has the same structure as in the $N=4$ case  (\ref{Hamiltonian_3}) \footnote{Now the fermions $\eta$ and $\psi$ change their roles as we use different parametrizations for the coset elements (\ref{Coset Element 4}) and (\ref{Coset Element N}). }, but now we have additional fermionic variables and the constraint (\ref{Constraint_Additional}). The analysis of the previous section indicates that there is only one first--class constraint associated to the broken $\kappa$--symmetry for $N>2$.
 It should be noted that if we analyzed the phase space of the models from previous sections without imposing the gauge fixing conditions, we would find the fermionic constraints of the similar structure. In particular, one would find that they have nonsingular massless limit $m\rightarrow0$. Due to the relations (\ref{Parameter restriction}) and (\ref{Restriction}) it implies that for massless particles $\kappa$--symmetry demands vanishing Wess--Zumino terms in (\ref{Action_S2}), (\ref{Action_S3}) and (\ref{Action_S4}).  In fact, classical Hamiltonian dynamics does not depend on the mass parameter at all, which can be seen by implementing the canonical transformation (\ref{Canonical redefinition})
 \bea
 &&
H=\frac{p^2}{2}+\frac{1}{2x^2}J^a J^a-\frac{i}{2x^2}(\eta\lambda^a\eta)J^a, \qquad p_\eta-\frac{i}{2}\eta=0,
\\\label{Constraint_Add}
&&
p_\psi+i\psi \left(\frac{p^2}{2}-\frac{2}{x^2}J^a J^a\right)-2p p_\eta+\frac{i}{x} J^a(\eta \lambda^a)=0,
\eea
that is nonsingular in the limit $m\rightarrow0$ as well. In the case $N=2$ $\k$--symmetry is unbroken and we have an additional first-class constraint that allows one to set $\psi=p_\psi=0$. In the general case the Dirac bracket constructed with respect to the fermionic constraints has a very complicated form.

\section{Conclusion}

To summarize, in this work we have constructed superconformal mechanics with the $OSp(N|2)$ supersymmetry group. First, we considered the instance of $N=4$. Two different coset spaces were considered, the corresponding bosonic parts being $AdS_2\times S^2$ and $AdS_2\times S^3$, on which the $\k$--symmetric actions have been constructed. Some of these models were shown to reproduce those previously known. In particular, the gauge--fixed version of the model (\ref{Action_S3}) on $AdS_2\times S^3$ is linked by a suitable canonical transformation to the work in \cite{Gal_D21}, while the model (\ref{Action_S2}) is exactly the $\k$--symmetric superparticle of \cite{Ch_2} in view of the isomorphism $osp(4|2)\simeq D(2,1;-\frac12)$. For arbitrary $N$ we have shown that the $\k$--symmetry is broken down to a one--parametric fermionic gauge symmetry, in agreement with a recent analysis of the $su(1,1|N)$ superparticle in \cite{Ch}. Interestingly enough, the Hamiltonian for this model has exactly the same structure as the one for the $N=4$ case. The additional fermionic variables, which are not gauged away due to the breakdown of the $\k$--symmetry, give rise to second--class constraints and modify the canonical bracket.
 We have also constructed an analogue of the model with semi--dynamical angular variables \cite{FIL,FIL_2}.

 Turning to possible future developments, it is of interest to generalize the classical treatment and consider our models at the quantum level. It is worth studying whether one can construct superparicle model with semi--dynamical variables in the superfield formalism. Another open problem is to deform this model in the spirit of \cite{Stepan_2} and to consider its quantum version \cite{Stepan}.
\vspace{0.5cm}

\section*{Acknowledgements}

\noindent
We thank A. Galajinsky for posing the problem and reading the manuscript.
This work was supported by the  the Foundation for Theoretical Physics Development “Basis”, RFBR grant 18-52-05002, the DAAD program and the Tomsk Polytechnic University competitiveness enhancement program.

\appendix
\numberwithin{equation}{section}

\section{MC one--forms for the case $N=4$}
As is known, the $\k$--symmetry reduces the number of fermionic degrees of freedom by half. For the case at hand, this is achieved by imposing the gauge fixing condition, setting the fermions corresponding to the generator $S$ to be vanishing. The gauge--fixed coset element reads
\be\label{Coset Element 4}
\tilde{G}=e^{tH}e^{zK}e^{\psi Q}u, \qquad u=e^{P^{1}_+\phi}e^{P^{2}_+(\t-\pi/2)}e^{P^{3}_+\chi}.
\ee
The corresponding bosonic MC one--forms are
\bea\label{MC forms_4}
&&
L_H=dt-i\psi d\psi, \qquad L_K=z^2dt+dz,\qquad L_D=2zdt,
 \nonumber\\[2pt]
&&
L^m=e^m+\frac{i}{2}(\psi\lambda^{p}_+\psi)O^{mp}L_K,
\eea
where
\be
u^{-1}J^m_+u=O^{mp} J^p_+,
\ee
and
\bea
&&
e^1=\sin\t\cos\chi d\phi+\sin\chi d\t, \qquad e^2=-\sin\t\sin\chi d\phi+\cos\chi d\t, \qquad e^3=-\cos\t d\phi+d\chi. \nonumber
\eea
The MC one--forms for the coset (\ref{Coset_2}) can be obtained from (\ref{MC forms_4}) by setting $\chi=0$.

\section{MC one--forms for the general case}
We choose a coset representative of (\ref{Coset_N}) in the form
\be\label{Coset Element N}
\tilde G=e^{tH}e^{\psi Q}e^{zK}e^{\eta S}u,
\ee
where $u$ is a coset element of $SO(N)/SO(N-1)$. The bosonic MC one–forms for this coset element read
\bea\label{MC forms N}
&&
L_H=dt-i\psi d\psi, \qquad L_D=2z L_H-2i (d\psi\eta),
\nonumber\\[2pt]
&&
L_K=z^2L_H+dz-2iz(d\psi\eta)-i\eta d\eta,
\nonumber\\[2pt]
&&
L^m=e^m+\frac{i}{2}(\eta\lambda^a\eta)O^{am}dt-(\eta \lambda^a d\psi)O^{am},
\eea
where $e^m$ are MC one forms on the coset $SO(N)/SO(N-1)$ coming from the term $u^{-1}du$. The matrices $O^{am}$ are given through the expression
\be
u^{-1}P^mu=O^{am}J^a.
\ee

\end{document}